\documentclass[aps,prb,twocolumn,superscriptaddress,showpacs]{revtex4}
\usepackage{graphics}
\usepackage{graphicx}
\usepackage{amsmath}
\usepackage{times}
\usepackage{bm}
\usepackage{amssymb}
\newcommand {\be} {\begin{equation}}
\newcommand {\bea} {\begin{eqnarray} \nonumber }
\newcommand {\ee} {\end{equation}}
\newcommand {\eea} {\end{eqnarray}}

\begin{document}
\title{Magnetic flux induced spin polarization in semiconductor multichannel rings
  with Rashba spin orbit coupling}
\author{G. S.  Lozano}
\email{lozano@df.uba.ar}
\affiliation{Departamento de F\'{\i}sica, FCEyN, Universidad de Buenos Aires,
 1428 Ciudad de Buenos Aires, Argentina }
\author{M. J. S\'anchez}
\email{majo@cab.cnea.gov.ar}
\affiliation{Centro At\'omico Bariloche, Comisi\'on
Nacional de Energ\'\i a At\'omica, 8400 San Carlos de Bariloche, Argentina.}

\begin{abstract}
We show that a finite magnetic flux threading a multichannel semiconductor
ring  induces spin accumulation at  the borders of the sample
when the Rashba spin-orbit interaction is taken into account 
even in the absence of external electric fields.
\end{abstract}
\maketitle
 
In recent years semiconductor devices in which the spin orbit (SO) interaction plays a significant role, have been proposed for applications in spin controlled transport \cite{spinbook}. Among them, multiple connected mesoscopic geometries are natural candidates to explore how spin dependent effects manifest in the quantum interference patterns that appear in  transport measurements at low temperatures \cite{morpu98,nitta99}.
In the case of quasi-1D rings, the effect  of the SO interaction  has been addressed theoretically in a series of papers \cite{meir89,loss92,balatsky93}.
As a consequence of the SO, the wave function acquires a non trivial spin dependent topological phase \cite{berry84} that manifests  in  several remarkable quantum phenomena.

We will be considering SO coupling of the Rashba type which arises on a 2DEG of a semiconductor heterostructure due to the inversion asymmetry of the confining potential \cite{rashba84}. 
When a ring is  pierced by a magnetic flux, the  SO modifies the magnetic flux dependence of the spectrum and therefore the conductance \cite{AronovG93} and the persistent current (PC), in the case of an isolated ring,  change as compared to the case without SO \cite{chaplik95}.
So far the  theoretical analysis has been restricted to 1D geometries  and quasi 1D in the two band approximation \cite{sgz02}, where the results are qualitative the same as in the 1D systems. 
The multichannel nature of realistic rings employed  in the experiments 
\cite{mailly93,meijer04} and recent controversies around  the possibility of experimentally observe a spin dependent phase in transport experiments with many propagating modes \cite{yau02,souma04} challenges us to  address  exactly the 2D geometry.

In this work we  show that  when a multichannel ring
with Rashba SO coupling  is pierced by a magnetic flux a spin accumulation effect is developed on the boundaries of the sample.
In addition, even for an even number of electrons, a finite spin polarization in the direction perpendicular to the plane of the ring is generated whose  intensity can be controlled with the magnetic flux.  
These phenomena share some analogies with the intrinsic Spin Hall Effect (SHE) studied in bar or T-like geometries \cite{kato04,Sinova04,wunder05}, but in the present  case the system is in equilibrium without external electric fields or voltage drops  applied \cite{usaj05}.

We start by considering a 2D  electron gas in the $xy$ plane confined to a
 mesoscopic annular region (multichannel ring) threaded by a magnetic flux
$\Phi$. 
The single particle Hamiltonian describing an electron of effective
mass $m$ subject to the Rashba SO coupling reads
\be
H=\frac{{\bf p}^2}{2m}+V+\frac{\alpha}{\hbar} ({\bf p} 
\times {\bf \hat{\sigma}}) \cdot \hat{z} \;,
\label{ham}
\ee
where $\alpha$ is the strength of the Rashba spin orbit (RSO) coupling 
 and the Pauli matrices ${\bf \hat{\sigma}}$ are defined as standard.
Employing polar coordinates  $\rho$ and $\varphi$ the hard wall
confining potential defining the ring is 
\be \label{pot} 
V(\rho)= 
\left\{
\begin{array} {l} 
0   \; \; \mbox{for   $r< \rho < R$}  \\ 
\infty  \; \; \mbox{otherwise} \; ,
\end{array} 
\right.  
\ee 
where $r$ and $R$ are the internal and external radii of the ring.
The vector potential 
which is introduced in the Hamiltonian via  the substitution, ${\bf p}= \hbar {\bf k}=-i \hbar{\bf \nabla} - 
\frac{e}{ c}{\bf A}$, is written in  the axial gauge as
${\bf A} = (\Phi / 2 \pi \rho) \hat {\varphi}$ .
Using , 
$\hat{\sigma}_{\rho}= \cos\varphi \hat{\sigma}_{x} + \sin\varphi \hat{\sigma}_{y}$ and
$\hat{\sigma}_{\varphi}= - \sin\varphi \hat{\sigma}_{x} + \cos\varphi \hat{\sigma}_{y}$ we 
can rewrite  the  Hamiltonian  as
\bea \label{h2d}
H=-\frac{\hbar^2}{2m}\left[ 
\frac{1}{\rho} \partial_{\rho}\left(\rho \partial_{\rho} \right)- 
\frac{1}{\rho^2}(i \partial_{\varphi} + \nu)^2\right] \\  
+ i  \alpha \hat{\sigma}_{\varphi} \partial_{\rho} - \frac{\alpha}{\rho} \hat{\sigma}_{\rho} 
\Bigl( i \partial_{\varphi} + \nu \Bigr) \, \, ,
\eea
where  $\nu=\frac{\Phi}{\Phi_0}$ is the magnetic 
flux in units of the flux quantum $\Phi_0= {h c}{/ e} $.
As $J_z=l_z+s_z=-i \hbar \partial_{\phi}+\frac{1}{2} \hbar \hat{\sigma}_z$, commutes with
$H$, the eigenfunctions can be chosen as 
\begin{equation}\label{sep} 
\psi_j(\rho,\varphi)= 
\left[ 
\begin{array}{c} 
e^{i(l)\varphi} \tilde{f}_l(\rho) \\ 
e^{i(l+1)\varphi} \tilde{g}_{l+1}(\rho) 
\end{array} 
\right]. 
\end{equation}
where $J_z \psi_j= \hbar j\psi_j$ and $j=l+\frac{1}{2}$.
In what follows it will be useful to work with dimensionless variables.
With that purpose   we define the dimensionless coordinate $\xi=\rho/R$, the aspect ratio $\lambda = R/r$ and 
\be
 \epsilon=\frac{2m E R^2}{\hbar^2}\equiv \frac{E}{E_0} \,\,,\,\, \beta=2R \frac{\alpha
   m}
{\hbar^2}
\,\, , \,\, f_l=R \tilde{f}_l \,\, , \,\, g_l=R \tilde{g}_l \; ,
\label{param}
\ee  
with the boundary conditions
\be
f_l({\lambda}^{-1})=g_{l+1}({\lambda}^{-1})=0= f_l(1)=g_{l+1}(1) \; .
\ee 
We can look solutions  of the form 
$f_l(\xi)\sim Y_{l-\nu} (k \xi)$ and $g_{l+1}(\xi)\sim Y_{l+1-\nu}(k \xi)$ 
 where $Y_l(\xi)$ are  Bessel functions of the type $J_l(\xi)$ or $N_l(\xi)$. 
The Rashba term  simply acts as rising or lowering  
operator on the Bessel function  basis since the following 
standard recurrence relations hold \cite{abram}: 
\begin{equation} 
\left(\frac{d}{d\xi}+\frac{1\pm l}{\xi}\right)Y_{l\pm 1}(k\xi) = \pm k Y_l(k\xi) \;.
\end{equation}
This is indeed the property which allows to obtain, as in the case of a disk geometry \cite{TLG}, an exact analytical 
solution. 
Due to the RSO, the bulk spectrum has two branches
\be 
\epsilon=k^2\pm \beta k\;. 
\ee 
Therefore for a given value of $\epsilon$ there are two non-trivial
solutions for the momentum $k$ that we denote $k^+$ and $k^-$ respectively. 
It is then possible to obtain  a solution as
\begin{equation}
Y(\xi)=\left(f_l\atop g_{l+1} \right)= \sum_{i=1}^{4} c_i Y_i(\xi)= \sum_{i=1}^{4} c_i \left(Y_i^1\atop Y_i^2 \right)
\end{equation}
with 
\bea
& Y_1( \xi)= \left(
\begin{array}{c}
J_{l-\nu}(k^+ \xi)\\ J_{l-\nu+1}(k^+ \xi) 
\end{array}
\right)
\;,
& Y_2( \xi)= \left(
\begin{array}{c}
J_{l-\nu}(k^- \xi)\\ -\frac{|\epsilon|}{\epsilon}J_{l-\nu+1}(k^- \xi) 
\end{array}
\right)
\eea
and with $Y_3$ and $Y_4$ obtained from $Y_1$ and $Y_2$ by exchanging Bessel
functions of type $J$ by Bessel functions of type $N$.

Defining $\tilde{Y}$ as,
\begin{equation}
 \tilde{Y} =\left(
\begin{array}{cccc}
Y_1^1({\lambda}^{-1}) & Y_2^1(\lambda^{-1}) &Y_3^1(\lambda^{-1}) & Y_4^1(\lambda^{-1})\\
 Y_1^2(\lambda^{-1})& Y_2^2(\lambda^{-1})& Y_3^2(\lambda^{-1}) &Y_4^2(\lambda^{-1})\\
 Y_1^1(1) & Y_2^1(1) & Y_3^1(1) & Y_4^1(1)\\
 Y_1^2(1) & Y_2^2(1)&  Y_3^2(1) & Y_4^2(1)\\
\end{array}
\! \right), 
\label{eigen}
\end {equation}
the boundary conditions lead to the equation $\det(\tilde{Y})=0$.
Given $\beta$, $\lambda$ and  $\nu$ we solve this equation
to obtain the (dimensionless) energies $\epsilon_{j,i} (\nu)$ where
$j$ is the total angular momentum and $i$  labels the different eigenstates
for a fixed $j$, in such a way that 
for $\nu =0$ and $\beta=0$ we have $\epsilon_{j,i} < \epsilon_{j,i+1}$.

In order to fix numerical estimates for  the parameters we consider
characteristic values extracted from recent experiments performed on
semiconductor heterostructures defined on a 2DEG. 
Rings with external radius $R \sim 400-500 nm$  and an aspect ratio  $\lambda \sim 2$ have been recently employed as devices \cite{meijer04}. Typical values for the Fermi wavelength are  $\lambda_F \sim 40-50 nm$ that give $k_F \sim 0.1 nm^{-1}$.
For  $R=400nm$ one gets  a maximum value of the      
(dimensionless) Fermi energy  $\epsilon_{F}= (R k_F)^2 \sim 1600$.
For an effective  mass $m \sim 0.042 m_e$, a Rashba coupling constant $\alpha= 8 meV nm$ and $R \sim 400$ we obtain $\beta = 2000 \; nm / R (nm) \sim 4$. These parameters define the sample {\bf S} studied in the present work.
As the relevant situation for an isolated system, we work in the canonical ensemble keeping fixed the total number of electrons $N$ as the magnetic flux is varied. One can  estimate $N$ at zero flux \cite{baltes}, that in this case gives $N = [ 3 \; \epsilon_F / 8 ] = 600$ (the symbol $[..]$ denotes integer part). 
The maximum number of transverse channels $M$ can  be  then calculated as \cite{fs00}
\be
M= \big[ \frac {2 \sqrt{N} (\lambda -1)}{\pi \sqrt{\lambda^2 -1}} \big] \;. 
\ee
Thus for $N \sim 200 -600$ one gets  $M \sim 4-8 $,  in agreement   with the
reported experimental values \cite{mailly93}.

For $\nu=0$  and finite $\beta$, the SO breaks the degeneracy between
states differing in one unit of $j$. The degeneracy between states with opposite values of $j$ is broken by the presence of a finite magnetic flux $\nu$,  being the  charge PC the signature of this broken symmetry.

As  discussed previously in the literature, 
in $1D$ rings the RSO induces  a topological phase \cite {AronovG93},
$ \Delta_{B} = \frac{\alpha \; m}{\hbar^2} \oint (\hat{z} \times \sigma)
\cdot d\mathbf{l} $, that  once added to the Bohm-Aharonov one, $2 \pi \nu$,  
leads to an "{\it effective flux}" $\nu_{eff}= \nu + \frac{1}{2}  (1  \mp \sqrt{(2\alpha \; m a /\hbar^2) +1})$ (the $\pm$ sign depends on the  sign of the $z$ spin projection in the {\it local spin frame} and $a$ is the radius of the ring). It is then via this effective flux that the SO interaction affects the behavior of the PC \cite{chaplik95}.

\begin{figure}[t]
\includegraphics[height=6cm,clip]{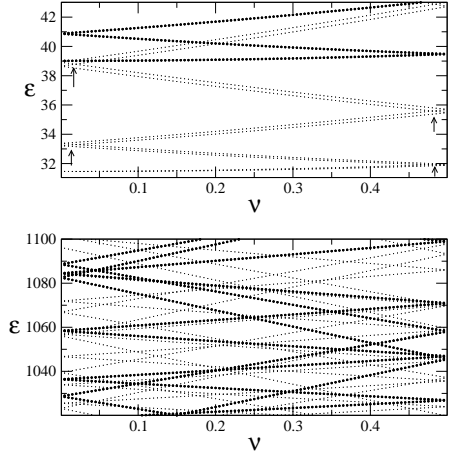}
\caption{Dimensionless energies $\epsilon$ as a function of the magnetic  flux 
$\nu$ for the  annular cavity $\bf{S}$ defined in the text.
Upper panel: lowest eigenvalues  in the absence of RSO (bold dotted lines)
and for $\beta=4$ (small dotted lines). The arrows are guides for the eyes to locate the position of the crossings due to the RSO coupling. See the text for details.
Lower panel: high  energy region showing many crossings between 
different transverse channels, in the absence of RSO (bold dotted lines)
and for $\beta=4$ (small dotted lines).}
\label{spec}
\end{figure}

The situation in the 2-D ring is considerably more involved. We display
in Fig.\ref{spec} some regions of the spectrum for both $\beta=0$ and $\beta=4$ cases. The upper panel of Fig.\ref{spec} shows the lowest eigenvalues of the  multichannel ring {\bf{S}} as a function of $\nu$.  As the first transverse channel is active in that region, the spectrum is similar  to that of the $1D$ ring (due to the symmetry respect to $\nu=0.5$ the spectrum is shown for $0 \le \nu \le 0.5$).
We can observe the evolution  with $\nu$  of the single particle energy states labeled by the quantum numbers $(j,i)$. For $\beta=0$ (bold dotted lines) we have the double degenerate fundamental states, $(\pm 1/2,0)$, and the doublets  $(1/2,1), (3/2,0)$ and $(-1/2,1), (-3/2,0)$. Notice that at $\nu=0$, except the fundamental state, the others  are four-fold degenerate and, as in the $1D$ regime, crossings occur only at  $\nu=0, 0.5$.
The effect of a finite RSO coupling is clearly visible in the figure. The RSO interaction lowers the energy of each state and it can even change the order in which they appear. For the case shown in Fig.\ref{spec}, the lowest energy states are $(3/2,0)$ and $(-3/2,0)$ that for finite flux remain almost degenerate and appear in the figure as a single line.
We then have four states which for very small $\nu$ (i.e, before any $\nu$ induced level crossing) are ordered as $(-1/2,0)$, $(1/2,0)$, $(5/2,0)$, $(-5/2,0)$. 
 Higher in energy, we display the states $(1/2,1)$, $(-1/2,1)$, $(-7/2,0)$, $(7/2,0)$.
Notice that as a result of the RSO new crossings appear. As an example,
we draw arrows in the panel as guides for the location of the new crossings between $(1/2,0)$ and $(5/2,0)$ and
$(-1/2,1)$, $(-7/2,0)$. These crossings are indeed the fingerprints of the effective flux $\nu_{eff}$ mentioned above \cite{note}.

When many transverse channels are  activated
the spectrum displays  additional crossings between levels belonging to different channels, even in the absence of RSO coupling (see lower panel of Fig.\ref{spec}). The crossings that arise due to the SO interaction are mixed with the crossings between levels with different transverse channel number and it is not straightforward  to identify the signature of an effective flux like in the 1D case, when only a single transverse channel is active. This can be understood looking at the functional form of the 2D Hamiltonian Eq.(\ref{h2d}), whose  last term contains the ratio $\alpha/ \rho$ between the RSO constant  and the radial coordinate. Therefore, loosely speaking, on average each transverse channel feels a topological phase that depends on the value of the transverse quantum number. 
This argument could be extended to explain  the difference in  patterns of conductance oscillations  of single-channel and multichannel open rings with RSO interaction \cite{souma04}.
In terms of the dimensionless variables, the only non vanishing component of the
charge current density for  eigenstates as given in  Eq.(\ref{sep}) reads,
\begin{equation}
J_{\varphi}=\frac{e\hbar }{m R^3\xi} \left((l-\nu) f_l^2 +
(l+1 -\nu) g_{l+1}^2 + \beta \; \xi \; f_l \; g_{l+1} \right) \;.
\end{equation}
Employing the  probability and spin densities, 
$\delta_j(\xi) \equiv f_l^2 + g_{l+1} ^2$, 
$\langle\hat{\sigma}_z\rangle_j\equiv \Psi^{\dagger}\hat{\sigma}_z\Psi$
and $\langle\hat{\sigma}_\rho \rangle_j\equiv \Psi^{\dagger}\hat{\sigma}_\rho \Psi$, the current can be written as 
\begin{figure}[t]
\includegraphics[height= 5cm,clip]{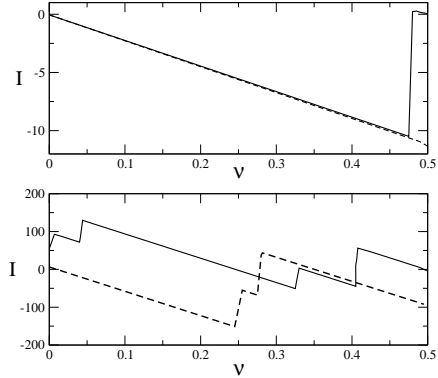}
\caption{ Persistent current $I$ as a function of  
$\nu$ in the absence of SO interaction (dashed lines) and for $\beta=4$ (solid line).
Upper panel corresponds to $N=6$ and lower panel to $N=182$. The finite slope of the jumps is due to size of the flux step ($5.10^-3$ ) employed in the numerical calculations }
\label{pcu}
\end{figure}
\begin{equation}
\label{current}
J_{\varphi}=\frac{e\hbar }{m R^3\xi} \left( ( j-\nu) \delta_{j}(\xi)
 -\frac{1}{2} \langle\hat{\sigma}_z\rangle_j+
 \frac{\beta}{2} \; \langle\hat{\sigma}_\rho \rangle_j \; \xi  \right) \;.
\end{equation}
Therefore  the effect of the SO interaction in  the charge current density is unveiled in the last 
term of Eq.(\ref{current}).   
To calculate the total charge PC we have to sum the contributions of all states up to the Fermi energy, 
\be
\tilde{I}=\int_r^R d\rho \sum J_{\varphi}^{(n)}(\rho)=-c \frac{\partial E}{\partial \phi}
\ee
where as before, $n$ labels the occupied states. Besides a geometrical factor that takes into account the area of the outer circle of the sample $\tilde{I}$ is the magnetic moment. In terms of the dimensionless variables,
\be
\tilde{I}=\frac{c E_0}{\phi_0} I \, ,\,\,\, I= - \frac{\partial \epsilon}{\partial \nu} \;,
\ee
For the parameters quoted before for sample $\bf S$, results  $\tilde{I} \sim 0.2  \; I (nA) $.
In Fig.\ref{pcu} we plot $I$ as a function of $\nu$ for $N=6 (M=1)$ and $N=182 (M=4)$ in order to show how
active open channels modify the behavior of the PC and the determination  of an effective flux as in 
the $1D$ regime \cite{note2}. 

As a result of the RSO interaction, spin projection is not a good quantum number. For  N occupied  states,  the mean value of the z-projection of the spin is proportional to
\be
\Sigma_z=\int_{ring}\sum_{n} \Psi_n^{\dagger}\hat{\sigma}_z\Psi_n= 
2 \pi\int_{\frac{1}{\lambda}}^1 \sigma_z(\xi) \xi d \xi
\ee
where $n=1,N(\epsilon_F)$ labels the occupied states and $\sigma_z (\xi)$ corresponds to the total  spin density for N particles.
Even in the case of N even, a non zero spin density is obtained  when SO is present  and $\nu \ne 0$.
In addition a spin accumulation effect is developed and it is manifested in the tendency of $\sigma_z$ to be positive on one border of the sample and  negative  on  the other one. 
Although the accumulation  becomes stronger as  N and the number of open channels increase, it is also present when only one transverse channel is active.
In order to explain the origin of the effect we first concentrate in a  pair of eigenstates that, being degenerate at $\nu =0$ with opposite value of $j$ and with opposite out of plane spin projection, become non-degenerated for finite $\nu$. The expectation value of $\hat{\sigma}_z$  in a given   $\Psi_j$ is $\langle\hat{\sigma}_z\rangle_j= \vert{f}_{l-\nu}|^2-\vert{g}_{l-\nu+1}\vert^2$. 
It is straightforward to verify that at  $\nu=0$ states with opposite value of $j$ have   opposite value of $\langle\hat{\sigma}_z\rangle$ and therefore for even number of particles $N=2p$ , is  ${\sigma}_z (\xi)=0$. 
As $\nu > 0$, two facts  induce a spin accumulation effect. On one hand,
the magnetic field breaks the symmetry between single particle states
with opposite value of $j$ and on the other hand, due to the presence of additional level crossings the total $J_z$ of the ground state can change and be different from zero for finite flux.
\begin{figure}[t]
\includegraphics[height=5cm,clip]{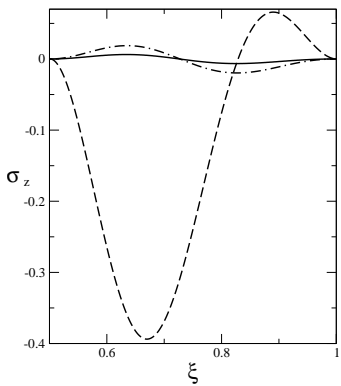}
\caption{Dimensionless total spin density $\sigma_z$  as a function of $\xi$  
for $\nu=0.2$ for the sample $\bf{S}$.
 $N=2$ (solid line), $N=4$ (dashed line) and $N=6$ (dotted-dashed line).}
\label{denmag1}
\end{figure}
In Fig. \ref{denmag1}, we show $\sigma_z$ for $\beta=4$, $
\nu=0.2$ and $N=2,4,6$. For $N=2$, the single particle states have $j=\pm 3/2$.
As $\nu$ is turned on, the effective orbital index $l_{eff} \equiv l-\nu$ becomes
$1-\nu$ and $-2-\nu$ for $j=3/2$ and $j=-3/2$ respectively. Thus as the modulus of the effective orbital index decreases (increases) for $j=3/2$ ($j=-3/2$), the probability density and $<\hat{\sigma}_z>_{j(-j)}$ are pushed toward the internal (external) boundary of the sample. This symmetry breaking explains the observed accumulation effect in this case. For $N=4$ the situation is different. Due to the level crossing mentioned before, the single particle states in the ground state
have $j=3/2,-3/2,-1/2,5/2$ and the accumulation effect is mainly due to the unbalance between the last two states. We show the case $N=6$ which again is similar to the case $N=2$.
 
\begin{figure}[t]
\includegraphics[height=5cm,clip]{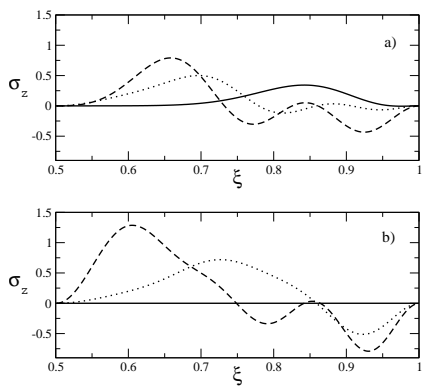}
\caption{Dimensionless total spin density $\sigma_z$  as a function of $\xi$  
for $\nu=0$ (solid line) , $\nu=0.1$ (dotted line)  and $\nu= 0.45$ (dashed line) for the sample $\bf{S}$.
a) $N=181$, b) $N=182$. Note that in this case is $\sigma_z=0$ for $\nu=0$ as expected for an even  number of particles.}
\label{denmag2}
\end{figure}

As the particle number is increased to the relevant experimental values, transverse channels are activated and the description of the effect becomes more complicated. Nevertheless  our simulations suggest that the accumulations effect is a generic feature of the system in the presence of a magnetic flux. 
As an illustration, in Fig. \ref{denmag2} we show the spin density $\sigma_z$ as a function of the dimensionless coordinate $\xi$ for $N=181$ and $N=182$  and different values of the magnetic flux in the range $0< \nu < 0.5$. For $N=181$ and $\nu=0$ (see Fig.\ref{denmag2} a)) a finite $\sigma_z(\xi)$ is obtained which value corresponds to the last occupied level, as expected for odd particle number.
On the other hand,  for $N=182$ is $\sigma_z(\xi)=0$ at  $\nu=0$, as it is shown in Fig.\ref{denmag2} b). 
For $\nu>0 $ the 
spin density profile has a more complicated structure than in the quasi 1D regime due to the
behavior  of the radial components of the spinors as the number of transverse channels increases and the flux is varied.

In summary we have shown that a finite magnetic flux threading a multichannel semiconductor 
ring with SO interaction  induces spin polarization with opposite sign for the two borders of the ring. We believe this system constitutes a new proposal to detect accumulation effects induce by SO interaction in constrain geometries in equilibrium, that is without applied electric fields or currents.
The characteristic wavelength of the accumulation effect  is of the order of $.1 \mu m$,  which is not far from the sensitivity of  methods employed
recently  to probe spin polarization in semiconductor channels \cite{kato04}. Besides the accumulation effect, the integrated  spin density $\Sigma_z $
is  different from zero and is sensitive to the value of the magnetic flux, as can be inferred from 
Fig.\ref{denmag2}. In the presence of an external magnetic field perturbation, the spin magnetization should be proportional to  $\Sigma_z $. With
the help of new experimental techniques based on resonant methods 
it should be possible to  sense changes in the total magnetization of isolated rings  due to SO interaction \cite{deblock02}. 

Partial financial support by ANPCyT Grant 03-11609, CONICET
and Foundaci\'on Antorchas Grant 14248/113 are gratefully acknowledged. We would like to thank C. Balseiro and G. Usaj for helpful discussions. G.S.L thanks ICTP, Trieste, where part of this work was done.


\begin{thebibliography}{27}
\bibitem{note}
Assuming a ring of radius $\hat R= (R + r)/2$ and employing the dispersion  relation $\epsilon (\nu_{eff})$ for 1D geometries \cite{AronovG93}, it is straightforward to obtain the values of $\nu_c$, given two levels with  $j-1$ and  $-j$ for example.
\bibitem{note2} 
In order to decrease CPU time in  the numerical simulations, and without loss of generality, we employ for  the present analysis these values of $N$.
Although they are smaller that the ones that correspond to the experimental reported $E_F$, the results and conclusions are completely general, as many transverse channels $(M \sim 4)$  are open.
\end{thebibliography}
\end{document}